\documentclass[conference]{IEEEtran}
\usepackage{upgreek}

\usepackage{url}

\usepackage{amsmath}

\usepackage[T1]{fontenc}

\usepackage{array}

\usepackage{graphicx}
\graphicspath{{./pdf/}{./jpeg/}{./images/}}
   \DeclareGraphicsExtensions{.pdf,.jpeg,.jpg,.png}

\ifCLASSINFOpdf
\else
\fi
\begin{document}
%
\title{BitTorrent Sync: Network Investigation Methodology}



%
\author{\IEEEauthorblockN{Mark Scanlon*,
Jason Farina*,
M-Tahar Kechadi}
\IEEEauthorblockA{UCD School of Computer Science and Informatics,\\
University College Dublin,
Belfield, Dublin 4, Ireland\\ Email: mark.scanlon@ucd.ie, jason.farina@ucdconnect.ie, tahar.kechadi@ucd.ie}}
%


\maketitle

\begin{abstract}
The volume of personal information and data most Internet users find themselves amassing is ever increasing and the fast pace of the modern world results in most requiring instant access to their files. Millions of these users turn to cloud based file synchronisation services, such as Dropbox, Microsoft Skydrive, Apple iCloud and Google Drive, to enable ``always-on'' access to their most up-to-date data from any computer or mobile device with an Internet connection. The prevalence of recent articles covering various invasion of privacy issues and data protection breaches in the media has caused many to review their online security practices with their personal information. To provide an alternative to cloud based file backup and synchronisation, BitTorrent Inc. released an alternative cloudless file backup and synchronisation service, named BitTorrent Sync to alpha testers in April 2013. BitTorrent Sync's popularity rose dramatically throughout 2013, reaching over two million active users by the end of the year. This paper outlines a number of scenarios where the network investigation of the service may prove invaluable as part of a digital forensic investigation. An investigation methodology is proposed outlining the required steps involved in retrieving digital evidence from the network and the results from a proof of concept investigation are presented.
\end{abstract}


%
\IEEEpeerreviewmaketitle


\section{Introduction}


Applications such as Evernote and Dropbox leverage the decreasing cost of hard disk storage seen in Infrastructure as a Service providers, e.g., Amazon S3, to provide data storage on the cloud to home users and businesses alike. The main advantage of services such as Dropbox, Google Drive, Microsoft Skydive and Apple iCloud to the end user is that their data is stored in a virtual extension of their local machine with no direct user interaction required after installation. It is also backed up by a fully distributed datacentre architecture that would be completely outside the financial reach of the average consumer. Their data is available anywhere with Internet access and is usually machine agnostic so the same data can be accessed on multiple devices without any need to re-format partitions or wasting space by creating multiple copies of the same file for each device. Some services such as Dropbox, also have offline client applications that allow for synchronisation of data to a local folder for offline access.

As Internet accessibility continues to become more commonplace and allows for increasingly faster access, it is not unexpected that many utilities that are intended for general use will be aided in the perpetration of some variety of cybercrime. One attribute that is highly desirable by those contemplating illegal activities is the notion of anonymity and data security, especially the ability to keep data secure transfer secure from inspection while in transit. BitTorrent Sync is a file replication utility that would seem to serve exactly this function for the user. Designed to be server agnostic, the protocol is built on already popular and widespread technologies that would not seem out of place in any network activity log. 

Each of the aforementioned services can be categorised as cloud synchronisation services. This means that while the data is synchronised between user machines, a copy of the data is also stored remotely in the cloud. In recent headline news, much of this data is easily available to governmental agencies without the need of a warrant or just cause. BitTorrent Sync (also referred to as BTSync, BitSync and bsync) provides the same synchronisation functionality (without the cloud storage aspect) and provides a similar level of data availability. The service has numerous desirable attributes for any Internet user \cite{bitsync}:

\subsubsection{Compatibility and Availability} Clients are built for most common desktop and mobile operating systems, e.g., Windows, Mac OS, Linux, BSD, Android and iOS.
\subsubsection{Synchronisation Options} Users can choose whether to sync their content over a local network or over the Internet to remote machines.
\subsubsection{No Limitations or Cost} Most cloud synchronisation services provide a free tier offering a small amount of storage and subsequently charge when the user outgrows the available space. BTSync eliminates these limitations and costs. The only limitation to the volume of storage and speed of the service is down to the limitations of the synchronised users machines.
\subsubsection{Automated Backup} Like most competing products, once the initial install and configuration is complete, the data contained within specified folders is automatically synchronised between machines.
\subsubsection{Decentralised Technology} All data transmission and synchronisation takes place solely in a Peer-to-Peer (P2P) fashion, based on the BitTorrent file sharing protocol.
\subsubsection{Encrypted Data Transmission} While synchronizing data between computers, the data is encrypted using RSA encryption. Under the BTSync API, developers can also enable remote file storage encryption \cite{bitsyncapi}. This could result in users storing their data on untrusted remote locations for the purposes of redundancy and secure remote backup.

As a result of the above desirable attributes, BTSync has grown to become a popular alternative to cloud synchronisation services. The technology had grown to over one million users by November 2013 and has doubled to two million users by December 2013 \cite{bitsyncstats}. The service will undoubtedly be of interest to both law enforcement officers and digital forensics investigators in future investigations. Like many other file distribution technologies, this interest may be centred around recovering evidence of the data itself, of the modification of the data or of where the data is synchronised to. 

While BTSync is based on the same technology as BitTorrent for the transfer of files, the intention of the application is quite different. This results in a change of users' behaviours, as well as a necessary change in the assumptions an investigator should make. BitTorrent is designed to be a one-to-many data dissemination utility. The uploader usually does not care about the identity of the downloader and a single seeder can deliver data to a large number of unique peers over the life of the torrent file. Data integrity and transfer speed take precedence over privacy of data in transit. 
BTSync on the other hand, is designed to be a secure data replication protocol for making a faithful replica of a data set on a remote machine. Data integrity is still highly prised but data privacy is now the top priority and speed-through-dispersion is sacrificed as a result. The files can only be read by users specifically given access to the repository. The advertisement of data availability is completely scalable by the owner with options ranging from restricting access to known IP addresses through to registration with a centralised tracker. Given the nature of the application, users are much more likely to know the operator of the remote site (this does not apply to secrets advertised online though that could be a point of commonality that would not necessarily have existed for pure BitTorrent clients).


\subsection{Aim and Contribution of this Work}

The aim of this work is to provide a reference for digital investigators discovering the use of BitTorrent Sync in an active investigation and for researchers working in the space. Providing the digital investigator with an explanation and methodology for investigating the service may aid in steering the investigation in a new direction. The contribution of this work firstly involves the formulation of a network investigation methodology for BitTorrent Sync, outlined in Section \ref{methodology}. This methodology includes recommendations for the investigation of a number of hypothetical scenarios where BTSync could be used to aid in criminal or illicit activities. While the legitimate usage of the system, e.g., backup and synchronisation, group modification, data transfer between systems, etc., may be of interest to an investigation, the technology may also be suitable in the aid of a number of potential scenarios of interest such as industrial espionage, copyright infringement, sharing of illicit images of children, etc., outlined in greater detail in Section \ref{btsync:usecase}. This work also documents each of the observed packets sent and received during regular operation of BTSync. Finally, the results from a proof-of-concept digital forensic investigation of the system are outlined in Section \ref{proofofconcept}.


\section{Background}
\label{background}


In order to gain an understanding of how BTSync functions, one must first understand the technologies upon which it is built. The application is a product built by BitTorrent Inc. (the creators and maintainers of the popular file-sharing protocol sharing the same time). As a result, the technologies used by the regular BitTorrent protocol and BTSync are developed using a similar premise. This section provides a brief overview of the required background information and outlined the key differences between the two applications.

\subsection{BitTorrent File Sharing Protocol}
\label{bittorrent}

The BitTorrent protocol is designed to easily facilitate the distribution of files to a large number of downloaders with minimal load on the original file source \cite{cohen2008bittorrent}. This is achieved through the downloaders uploading their completed parts of the entire file to other downloaders. A BitTorrent swarm is made up of both seeders (peers with complete copies of the content shared in the swarm), and leechers (peers who are downloading the content and may have none or some of the content). Due to BitTorrent's ease of use and minimal bandwidth requirements, it lends itself as an ideal platform for the unauthorised distribution of copyrighted material. The unauthorised distribution of copyrighted material typically commences with a single original source sharing large sized files to many downloaders.

\subsubsection{Bencoding}
\label{bencoding}
Bencoding is a method of notation for storing data in an array list. The main advantage of bencoding is that it avoids the pitfalls of system-byte order requirements (such as big-endian or little-endian), which can cause issues for cross platform communication between applications. The datagram packet can easily be converted to a human readable UTF-8 encoded sequence of \texttt{key:value} pairs. Indicative \texttt{key:value} pairings are presented in Table \ref{tab:bencoding}.

\begin{table}
\caption{BTSync Packet Bencoding Fields}

\makebox[\linewidth]{
    \begin{tabular}{|l|p{2.9in}|}
\hline
\textbf{Key} & \textbf{Explanation}  \\ \hline
	d:      & Marks the start of a dictionary     \\ \hline
	l:      & List start, the start of a list of field:value pairs in an array. Lists are terminated with an ``e'' \\ \hline
	la:      & local Address IP:Port in Network-Byte Order      \\ \hline
	ea:     & External Address IP:Port in Network-Byte Order\\ \hline
	m:      & Message Type Header, e.g., ping      \\ \hline
	peer:      & [Peer ID]      \\ \hline
	share:      & [Share ID]      \\ \hline
	nonce:    &  16-byte nonce for key exchange between peers negotiating data exchange \\ \hline
	e:      & Marks the end of a dictionary or list      \\ \hline
    \end{tabular}
}
\label{tab:bencoding}
\end{table}

The value for any pair is stored as a sequence of-bytes with the exception of integer values. Associated with the integer indicating keys, bencoding uses the lowercase ``i'' to indicate the start of an integer value, which is also terminated with a lowercase ``e''.

\subsubsection{Active Peer Discovery}
\label{discovery}

Each BitTorrent client must be able to identify a list of active peers in the same swarm who have at least one piece of the content and is willing to share it, i.e., identify a peer that has an available open connection and has the bandwidth available to upload. By the nature of the implementation of the protocol, any peer that wishes to partake in a swarm must be able to communicate and share files with other active peers. BitTorrent provides a number of methods available for peer discovery. There are a number of methods that a BitTorrent client can use in an attempt to discover new peers who are in the swarm outlined below

\begin{enumerate}
\item Tracker Communication -- BitTorrent trackers maintain a list of seeders and leechers for each BitTorrent swarm they are currently supporting \cite{cohen2003incentives}. Each BitTorrent client will contact the tracker intermittently throughout the download of a particular piece of content to report that they are still alive on the network and to download a short list of new peers on the network.
\item Peer Exchange (PEX) --  As set out in the standard BitTorrent specification, there is no intercommunication between peers of different BitTorrent swarms besides data transmission. Peer Exchange is a BitTorrent Enhancement Proposal (BEP) whereby when two peers are communicating (sharing the data referenced by a torrent file), a subset of their respective peer lists are shared during the communication.
\item Distributed Hash Tables (DHT) -- Many BitTorrent clients, such as Vuze and $\upmu$Torrent contain implementations of a common distributed hash table as part of the standard client features. The common DHT maintains a list of each active peer using the corresponding clients and enables cross-swarm communication between peers. Each known peer active in swarms with DHT contributors is added to the DHT. The mainline BitTorrent DHT protocol (also used by BTSync), is based on the Kademlia protocol. Regular BitTorrent file-sharing users and BTSync users contribute to the update and maintenance of the DHT. 
The DHT provides an entirely decentralised approach aiding in the discovery of new peers sharing particular pieces of content. The Kademlia DHT structures its ID space as a tree \cite{li2005comparing}. The distance between two keys in the ID space is their ``exclusive or'' (\texttt{xor}). Each user in the DHT generates a unique key that is used for identification when connecting to the DHT. The piece of the DHT that each peer stores is related to this \texttt{xor} calculation, i.e., those peer IDs that are closest to the key (e.g. a torrent's \texttt{info\_hash}) are responsible for facilitating lookups for those keys. The same DHT responsible for regular BitTorrent file-sharing is also responsible for maintaining a lookup for BTSync shared content. In this scenario, the key used is based on the public read-only key generated for each shared folder in BTSync.

While a DHTs decentralised nature results in a much more resilient service compared to server based tracker, it also results in it be vulnerable to certain attacks, as outlined in greater details in Sit et. al's 2002 paper \cite{sit2002security}.

\item Local Peer Discovery (LPD) -- This is enabled by checking the ``Search LAN'' option in most BitTorrent client's application preferences. When enabled the application will announce its availability to potential local peers using multicast packets. Once a client on the network receives a multicast packet, that client will check its current list of shares to see if a match is found. Is a match it found, that peer will respond to the origin of the request offering to synchronise the content.
\end{enumerate}

%

\subsubsection{Downloading of Content through BitTorrent}
\label{downloading}

To commence the download of the content in a particular BitTorrent swarm, a metadata \texttt{.torrent} file or a corresponding magnet universal resource identifier (URI) must be acquired from a BitTorrent indexing website. This file/URI is then opened using a BitTorrent client, which proceeds to identify other active peers sharing the specific content required. The client application then attempts to connect to several active members and downloads the content piece by piece. Each BitTorrent swarm is built around a single piece of content which is determined through a unique identifier based on a SHA-1 hash of the file information contained in this UTF-8 encoded metadata file/URI, e.g., name, piece length, piece hash values, length and path.

\subsection{BitTorrent Sync}
\label{bysync}



BTSync is a file replication utility created by BitTorrent Inc. and released as a private alpha in April 2013 \cite{bitsync}. It is not a cloud backup solution, nor necessarily intended as any form of off-site storage. Any data transferred using BTSync resides in whole files on at least one of the synchronised devices.  This makes the detection of data much simpler for digital forensic  purposes as there is no distributed file system,  redundant data block algorithms or need to contact a cloud storage provider to get a list of all traffic to or from a container using discovered credentials. The investigation remains an examination of the local suspect machine. However, because BTSync uses DHT to transfer data there is also no central authority to manage authentication or log data access attempts. A suspect file found on a system may have been downloaded from one or from 1000 sources and may have been uploaded to one or more recipients. Additionally while the paid services offer up to 1TB of storage (Amazon S3 paid storage plan) the free versions which are much more popular with home users cap at approximately 10GB. BTSync is limited only by the size of the folder being set as a share.  Another concern for any investigation into BTSync folders is that unless the system being examined is the owner/originator of the folder being shared, it is quite possible that any files present were downloaded without prior knowledge of their content or nature. BTSync has no built in content preview facility in its protocol, it blindly synchronises from host to target without any selection process available to the user. In fact, if the user were to delete a file from a read only share, that file would be re-created the next time the folder synchronised with the parent folder as long as the file was still present on the original location.

\subsubsection{Secrets}
\label{secrets}

\begin{figure}
\centering
\includegraphics[width=0.25\textwidth]{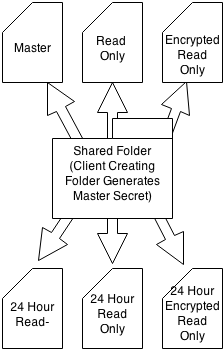}
\caption{Versions of Secrets Available for each Shared Folder}
\label{fig:secrets}
\end{figure}

Secrets are the unique identifiers used by the BTSync service to differentiate between shared folders. In order for the 20-byte secrets to be human readable, they are displayed using Base32 encoding \cite{bitsync}. BTSync facilitates the generation of three categories of secrets for the sharing of data contained within specific folders, as can be seen in Figure \ref{fig:secrets}:

The first is a read/write or master secret. This secret is a randomly generated key created by the client application on the machine initialising any given shared folder (this is created using \texttt{/dev/random} on *nix systems or \texttt{CryptoApi} on Windows systems). Anyone with access to this secret is granted both read and write access to the shared folder. For example, each user using this secret is able to add files to be replicated across all other member machines.  Once this access is granted for any client, it cannot be revoked without the creation of an entirely new share (and thus starting afresh with shared nodes). However, once a machine has download the shared content, there is no method available to reliably remotely delete any content. By default, if content is remotely deleted, this change will be reflected on each member machine but the delete file will be copied into the \texttt{.SyncArchive} folder and stored for 30 days before deletion. Depending on the version of BTSync used to create the initial shared folder, this 20-byte generated secret will be prepended with the character `A' or `D' to form a 42-byte secret for distribution. As an added layer of security, a user can opt not to use a 20-byte secret generated by BTSync and can instead use their own base64 encoded secret that is at least 40 characters in length. This allows users to include both upper and lower case characters in their key (while the default BTSync secrets are all uppercase).

The second is a read-only secret. This secret allows remote hosts to partake in a one-way synchronisation of a folder from those with read-write permissions. The 20-byte generated secret is prepended with the character `B' or `E', depending on version used.

The final option is an encrypted read-only secret. This secret will enable remote machines to synchronise an encrypted copy of the shared content. The remote machine only has read access to the share. Using these encrypted read-only keys, it is possible for two or more peers to agree on a reciprocal remote backup agreement without ever needing to expose the content of the folders being synced between them. This encryption secret can only be generated with more recent version of BTSync and will begin with the character `F'.

For each of the secret categories outlined above, the generated secret will always map to that specific shared folder of content with the specified level of access. However, the user can also create a time sensitive variation for each of those secret, as shown towards the bottom of Figure \ref{fig:secrets}. This variation involves specifying that joining that shared folder will only be permissible for a period of 24 hours. After the 24 hour window, the secret (irrespective of access type) will become invalid but users who have already added this secret to their BTSync installation will continue to have access to updates from shared folder (and optionally the continued ability to add content to the folder). Once added the secret recorded in \texttt{sync.dat} will be the master secret (R/W) or read-only secret, not the 24 hour secret received.

Each of the secret types outlined above need never necessarily be shared publicly, i.e., any user can create a number of secrets solely for his personal use across his different machines. Depending on the level of access the user wishes to give to a third-party, he can give the corresponding secret to any other user through regular one-to-one communication methods (e-mail, instant messaging, social networking, SMS, etc.). If public distribution is desirable, there are a number of public online avenues for BTSync users to share secrets with each other (e.g. \texttt{www.btsynckeys.com}, \texttt{http://www.reddit.com/r/btsecrets}, among others.)

\subsection{Potential Scenarios Pertinent to Digital Forensic Investigation}
\label{btsync:usecase}


\subsubsection{Industrial Espionage} Many companies are aware of the dangers of allowing BitTorrent traffic on their networks. However, quite often corporate IT departments enforce a blocking of the technology through protocol blocking rules on their perimeter firewalls. This has the effect of cutting off any BitTorrent clients installed on the LAN from the outside world. In addition to Deep Packet Inspection (DPI) to investigate the data portion of a network packet passing the inspection point, basic blocking of known torrent tracker sites using firewall rulesets can be used. BTSync does use BitTorrent as the protocol for file transfer but once the transfer session is established using the BTSync protocol all traffic is encrypted using AES and may not be open to inspection by a firewall. It also does not follow the current known patterns that would identify an encrypted BitTorrent stream as the target-source profile is different. Blocking \texttt{t.usyncapp.com}  and \texttt{r.usyncapp.com} will stop the tracker and relay options from being used but BTSync can operate quite well without those services. Local peer discovery can use multicast or direct "known peer" configuration where a known \texttt{IP:Port} combination is used to identify a specific machine allowed to participate in the share. This specificity would negate the issue of multicast packets usually not being routed beyond the current network segment. A scenario where BTSync can be used to transfer files within a LAN would be to transfer data to a machine with lower security protocols in place such as the capability to write to a USB device or perhaps even unmonitored access to the Internet (and the BitTorrent protocol ) through a designated guest LAN.

\subsubsection{Cloudless Backup} By synchronizing between two or more machines accessible to the user, data can be stored in multiple locations as a form of backup. The secondary copies of a file would be stored using a read only key so that only changes on the primary system will ever replicated.
A feature of BTSync that is enabled by default but can be disabled in the configuration file, is the use of the \texttt{.SyncArchive} folder that stores a copy of any file deleted or changed for a preset period of time allowing for a form of file recovery or versioning.

\subsubsection{Encrypted Remote P2P Backup} The BitTorrent Sync API \cite{bitsyncapi} adds the functionality to generate an ``encryption secret''. Through the use of encryption secrets, a BTSync user has the ability to remotely store encrypted data, e.g., personal, sensitive or illegal, on one or more remote machines. These remote machines do not have the ability to decrypt the information stored. The data could then be securely wiped off the original machine and easily recovered at a later stage.

\subsubsection{Dead Drop} Due to BTSync's intended use as a file replication utility, it is assumed that a person receiving a copy of a shared directory is aware of the contents of the folder. As a result, no method was included to gather details of the contents of a share before synchronisation. The API \cite{bitsyncapi} introduced this function but only a node configured correctly with an API key will return a folder or file listing when queried. 

\subsubsection{Secure P2P Messaging} For example, the proof of concept found at \texttt{http://missiv.es/}. The application currently operates by saving messages to an ``outbox'' folder that has a read only key shared to the person you want to receive the message. They in turn send you a read only key to their outbox. One to many can be achieved by sharing the read only key with more than one person but no testing has been done with synchronisation timing issues yet and key management may become an issue as a new outbox would be needed for each private conversation required.

\subsubsection{Piracy} -- BitTorrent, like any other P2P technology, was designed for one-to-many distribution of large content and has become almost synonymous with piracy. BTSync was not necessarily intended to be a one-to-many distribution utility. However, it does allow for a group of users to set one another as ``known peers'' so that they can communicate directly through encrypted channels. Websites such as \texttt{http://bitsynckeys.com/} have examples of users posting keys publicly and advertising the content as being copyrighted material.

\subsubsection{Serverless Website Hosting} -- This involves the creation of static websites served through a BTSync shared folder. These websites could be directly viewed on each user's local machine. The local copies of the website could receive updates from the webmaster automatically through the synchronisation of the content associated with a read only secret.

\subsubsection{Malicious Software Distribution} -- Due to the lack of any trust level being associated with any publicly shared secret, the synchronised files may contain infected executables.

For each of the above scenarios, an added dimension can be created by the BTSync user: time. Due to the ability to create ``throw away'' or temporary secrets for any piece of content, the timeframe where evidence may be recovered from remote sharing peers might be very short.

\section{Related Work}
\label{related}

This paper is focused on the network communication protocol employed by BTSync and the investigation thereof. The work presented as part of this paper builds upon the work of Farina et al. \cite{farina2014}, which outlines the forensic analysis of the BTSync client application on a host machine. This paper outlines the procedures for identifying a current or previous install of the BTSync application and the extraction of secrets from gain physical access to a machines hard drive and performing a regular digital forensic investigation on its image. At the time of publication, there are no other academic publications focusing on BTSync. However, seeing as BTSync shares a number of attributes and functionalities with cloud synchronisation services, e.g., Dropbox, Google Drive, etc., and is largely based on the BitTorrent protocol, this section outlines a number of related case studies and investigative techniques for these technologies.

\subsection{BitTorrent Forensics}
\label{btforensics}
Numerous investigations have been made into identifying the peer information of those involved in BitTorrent swarms. Most of these publications focus on the investigation of the unauthorised distributed of copyrighted material \cite{layton2010investigation}, \cite{scanlon2010week} and \cite{le2010spying}. Depending on the focus of the investigation, peer information may be recorded for a particular piece of material under investigation or a larger landscape view of the peer activity across numerous pieces of content.

\subsection{Client-side synchronisation Tool Forensics}
\label{clientside}

Forensics of cloud storage utilities can prove challenging, as presented by Chung et al. in their 2012 paper \cite{Chung201281}. The difficulty arises because, unless complete local synchronisation has been performed, the data can be stored across various distributed locations. For example, it may only reside in temporary local files, volatile storage (such as the system's RAM) or dispersed across multiple datacentres of the service provider's cloud storage facility. Any digital forensic examination of these systems must pay particular attention to the method of access, usually the Internet browser connecting to the service provider's storage access page (https://www.dropbox.com/login for Dropbox for example). This temporary access serves to highlight the importance of live forensic techniques when investigating a suspect machine as a ``pull out the plug'' anti-forensic technique would not only lose access to any currently opened documents but may also lose any currently stored sessions or other authentication tokens that are stored in RAM.  


In 2013, Martini and Choo published the results of a cloud storage forensics investigation on the ownCloud service from both the perspective of the client and the server elements of the service \cite{Martini2013287}. They found that artefacts were found on both the client machine and on the server facilitating the identification of files stored by different users. The module client application was found to store authentication and file metadata relating to files stored on the device itself and on files only stored on the server. Using the client artefacts, the authors were able to decrypt the associated files stored on the server instance.
%
%
%

\section{BitTorrent Sync Network Protocol Analysis}
\label{protocol}

Unless configured otherwise through application options or the configuration files, BTSync will attempt to contact the server at \texttt{t.usyncapp.com}. The DNS request resolves to three IP addresses: \texttt{54.225.100.8}, \texttt{54.225.92.50} and \texttt{54.225.196.38}. These three IP addresses are servers hosted on Amazon's EC2 cloud service. This is the BTSync tracker server, which facilitates peer discovery for clients looking to synchronise data. One peer request message is sent for each share stored on the local machine and the act of requesting a peer lookup also serves to register the requesting client as a source for that share. 

Packets sent from the client to the tracker server contain registration details and \texttt{get\_peers} message requests  (when a new share is created it registers the share with the tracker using a \texttt{get\_peers} packet). A \texttt{get\_peers} packet takes the form of:\newpage
\noindent \texttt{BSYNC[00] \\d2:la6:[6-byte IP/Port] \\1:m9:get\_peers \\4:peer20:[20-byte peerID] \\5:share20:[20-byte ShareID] e} \\ \small (where the observed keys are defined in Table \ref{tab:tpacket})
\normalsize

\begin{table}
\caption{Sample Tracker Packet}
\begin{tabular}{|l|p{2.9in}|}
\hline
BSYNC  & The Header that signifies the start of BTSync data  \\ \hline
0x00   & Null \\ \hline
d & Start of the dictionary of key:value pairs   \\ \hline
2:la  & Local Address Label identifier which consists of 6-bytes, the first 4 are IP, the last two are Port   \\ \hline
2:ea  & 6-byte External IP:Port pair \\ \hline
6:    & Local IP:Port pair \\ \hline
1:m   & Message label identifier \\ \hline
x:    & x Length message type value \\ \hline
4:peer  & Local peer label \\ \hline
20:   & Local PeerID  \\ \hline
5:share & Local ShareID label \\ \hline
20:    & The 20 character ShareID is the SHA1 of the secret used and can be found in the \texttt{.SyncID} file contained in the share as well as the name of the corresponding SQLite 3 database file in the \texttt{.Sync} folder.\\ \hline
\end{tabular}

\label{tab:tpacket}
\end{table}

This packet is sent to the tracker server once a second and a separate packet is sent for each share present on the local machine. It is noteworthy that, even when a new share is created, the first packet advertising that share to the server uses a message type of get\_peers. 

The receiving server will respond to the Client with a similar packet in the form:\\
\noindent \texttt{BSYNC[00] \\d2:ea6:[external IP:Port of the requesting peer] \\1:m:5:peers[PEER LIST] \\5:share20:[20-byte shareID] \\4:time:i[unix timestamp] e }
\normalsize

The response containing the peer list will always contain at least the originating peer information and as a result will never be empty. The peer list takes the form of:\\
\noindent \texttt{1:l \\d1:a6:[6-byte external IP:Port] \\2:la6:[6-byte local IP:Port] \\1:p20:[20-byte PeerID] e}

For each peer that has contacted the tracker advertising the relevant ShareID will be included as an entry in the list returned following the format above.

\subsection{Local Peer Discovery} 
When the option to search LAN is enabled the application will start sending out multicast packets to port 3838 across the LAN. The multicast packets are BTSync bencoded packets with the following format and the keys are further explained in Table \ref{tab:mcastping}:\\
\noindent \texttt{BSYNC[00] \\d1:m4:ping4:peer20:[20-byte Peer ID] \\4:port[i Integer e] \\5:share32:[32-byte content ShareID] e}\\
\normalsize

\begin{table}
\caption{Multicast Ping Packet}
\begin{tabular}{|l|p{2.9in}|}
\hline
BSYNC  & The BTSync Header  \\ \hline
0x00   & Null \\ \hline
d & Start of the dictionary of key:value pairs   \\ \hline
1:m   & Message label identifier \\ \hline
4:PING    &  The message type \\ \hline
4:peer  & Local peer label \\ \hline
20:   & PeerID of the multicasting Peer  \\ \hline
5:share & Local ShareID label \\ \hline
32:    & Hash function of the secret found in the \texttt{.SyncID} file (This also corresponds to the SQLite 3 database filename in the \texttt{.Sync} folder) \\ \hline
\end{tabular}

\label{tab:mcastping}
\end{table}
Once a peer receives a multicast message that contains a ShareID that it possesses the peer responds with the content: \\\\
\noindent \texttt{BSYNC[00] \\d1:m4:ping4:peer20:[20-byte PeerID] \\4:port[i Integer e] \\5:share20:[20-byte ShareID] e}\\
\normalsize
The keys have the same definitions as those shown in Table \ref{tab:mcastping} with the only exception being that the ShareID does not have the additional 12-bytes present in the received packet and that the ShareID is the ShareID local to the responding machine and stored in the local \texttt{.SyncID} file.

\subsection{BTSync Relay Server}
\label{BTSyncreq}

When BTSync finds that it needs to communicate directly between two firewalled peers, the application may make use of a relay server. This relay server option is available if the ``Use Relay Server if available'' option is enabled in the configuration. The relay server is contacted by a DNS request sent out for \texttt{r.usyncapp.com}, which resolves to the following IP addresses: \texttt{67.215.229.106} and \texttt{67.215.231.242}.

These are the IP addresses of the relay servers contactable on remote port 3000. Each peer contacts the relay server using an outbound connection that should bypass any firewall rule preventing unauthorised inbound connections. Once the server handshake has taken place, the negotiation to set up a secure connection between the two peers begins. The following sequence of events is observable:

\subsubsection{}  Peer contacts the relay server to initiate contact with the remote peer.\\
\noindent \texttt{BSYNC[80][20-byte remote peerID]\\d1:m4:ping4:peer20:[20-byte local peerID]\\5:share32:[32-byte shareID] e}
\normalsize

\subsubsection{}  The relay server responds to the peer using the remote server peer ID as the message header.\\
\noindent \texttt{BSYNC[80][remote peerID]\\d1:m4:ping4:peer20:[remote peer ID]\\5:share32:[32-byte ShareID] e}
\normalsize

\subsubsection{}  The peer contacts the relay with a hashmap of the share to indicate which parts are required.\\
\noindent \texttt{BSYNC[80][remote peerID][non-bencoded data including a hashmap] e}
\normalsize

\subsubsection{}  The server responds to the peer with a hashmap of the remote share to conclude the exchange of data availability.\\
\noindent \texttt{BSYNC[80][remote peerID][non-bencoded data including hashmap] e}
\normalsize

\subsubsection{}  The peer contacts the server to arrange transfer of the data and to supply the nonce for encrypted traffic and provide a status ID.\\\\
\noindent \texttt{BSYNC[80][remote peerID][non-bencoded data including hashmap request]\\d5:nonce16:[nonce]5:share32:[share ID]\\3:sid16:[sid value] e}
\normalsize

\subsubsection{} The relay contacts the peer to confirm file part availability\\
\noindent \texttt{BSYNC[80][remote peerID][non-bencoded data] e}
\normalsize

\subsubsection{} The relay server Confirms the SID status and supplies the remote nonce to complete the bridge for encrypted data transfer\\
\noindent \texttt{BSYNC[80][remote peerID]d5:nonce16:[nonce data]3:sid2:OK e }
\normalsize

\subsubsection{} Encrypted bidirectional traffic transfer commences with the relay server acting as the router delivering packets to each peer.

\subsection{BTSync Data Transfer}
\label{InternetSync}

The transfer of data during a BTSync synchronisation operates in a similar fashion as a regular BitTorrent download as described in Section \ref{downloading} above. A unique magnet URI is created for each file contained within the shared folder and this is used for requesting chunks of the entire file from known peers sharing this content.

\subsection{Differentiation from Regular BitTorrent Traffic}
\label{differentiation}

While much of the network topology of BTSync is shared with regular BitTorrent, the request and response packets differ from those employed by regular BitTorrent file-sharing traffic. The most obvious addition is the \texttt{BSYNC} header attached to each datagram transmitted on the network. Besides that addition, the active peer list that is returned also contains additional information over the regular BitTorrent file-sharing protocol: namely the inclusion of the local IP:port address pairs for each peer. From an investigative perspective, this extra information could prove useful in identifying the particular machine involved in the BTSync network as opposed to merely resolving the WAN IP address back to a router with potentially hundreds of LAN users. The local DHCP records could be used to resolve the MAC address (and often the hostname) of the individual machine identified during the network investigation.

In addition to the regular BitTorrent peer discovery methods outline in Section \ref{discovery} above, BTSync also allows the user to manually add known IP addresses to the local cache of peers. BTSync facilitates this through the option to add ``Predefined Hosts'' to the configuration or application options. These are hardcoded IP address and port entries that are saved in order of preference. BTSync will contact these peers directly, without any requirement for a multicast (LPD) or sending a get\_peer request to an online tracker.

\section{Investigation Methodology}
\label{methodology}

\begin{figure}
\centering
\includegraphics[width=0.3\textwidth]{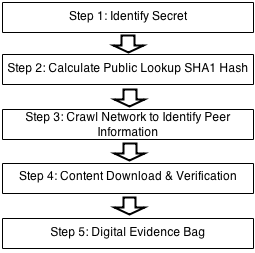}
\caption{Steps Involved in Performing a BTSync Network Investigation}
\label{fig:methodology}
\end{figure}

This section outlines a reproducible methodology for the network investigation methodology. Depending on which of the scenarios outlined above, the methodology may branch according to what the desired outcome will be. Figure \ref{fig:methodology} outlines the five steps involved in the investigative process (each of these steps are described in greater detail below).

\subsection{Identification of Content}
\label{identification}

Depending on the scenario that motivates the BTSync network investigation, there are a number of avenues that the forensic investigator may find secrets (and corresponding hash values) needed for investigation:
\subsubsection{Web Discovery} -- As soon as BTSync was released as a public alpha, publicly accessible sharing secrets started to appear online. Two ``subreddits'' appeared on Reddit \cite{reddit} and numerous websites and blogs were created to set up an online ``dead drop'' secret share, for example \texttt{http://www.12char.com} and \texttt{http://www.btsynckeys.com}. It is also feasible that an investigator could come across an online community that shares secrets in a private forum for the purposes of trading data and material without 3rd party involvement.

\subsubsection{Local Discovery} -- An investigator could, in the course of an investigation find evidence of BTSync having been used to transfer material to the suspect machine. This could be that BTSync installed and the folder listed in the list of shares stored in the configuration file , webUI or the BTSync hidden \texttt{.Sync} folder. BTSync log files (/.sync/sync.log) , or, if BTSync is not present (uninstalled) there could still be \texttt{.SyncID} files remaining in folders that were synchronised from remote peers. A hexdump of the \texttt{.SyncID} file or, more conveniently, the names of the db files found in the \texttt{.Sync} folder will give the SHA1 encoded share ID that the investigator needs to find other peers actively sharing that content

\subsubsection{LAN traffic} -- Many companies configure their edge firewalls to block torrent traffic for the general users. If the company uses torrent for some other business purpose it will usually be accounted for and allowed from or to a particular server or subnet. However, BTSync allows for all external communicate beyond the LAN to be turned off (in the configuration file or in the settings dialogue the options for ``Use DHT'', ``Use Tracker'' and ``Use Relay Server'' can be disabled) leaving only the settings for LAN discovery or known peers. A security review of the router logs may find active torrent traffic within the LAN or system admins may discover evidence of torrent applications run.

\subsection{Identification of Lookup Hash}
\label{hash}
Requesting a list of peers through any of the peer discovery methods outlined above requires a unique lookup hash. This hash is used by the tracker, DHT, PEX and LPD in the association of know peers to a particular piece of content.

\subsection{Crawl the Network to Identify Peer Information}
\label{crawl}
Each of the peer discovery methods outlined above should be queried for a list of known active nodes sharing that content. Due to the user configurable nature over which services are enabled in the BTSync client, to ensure complete node enumeration/identification, the results from each of the peer discovery methods should be combined to form the final result of collected information.

\subsection{Downloading and Verification of Content}
\label{verification}
Depending on the scenario being investigated, it may be necessary to download a copy of the content stored remotely for investigation or verification. In order to accomplish this, a regular BitTorrent download can be started for each of the files contained within the shared folder. If the investigation's goal is to attempt to recreate content deliberately deleted off a suspect's machine, the data can only be entirely recovered if there is a complete copy of the data stored remotely. However, this does not mean that any single node needs to have 100\% of the content. The original data can be recombined so long as a complete copy exists split among the distributed nodes actively sharing the content.

\subsection{Digital Evidence Bag}
\label{DEB}
Once the required information is gathered, the resulting data and all associated metadata (peer information, file sizes, hash values, etc.) should be gathered together into a suitable digital evidence bag. For verifiable reproduction of the results achieved, a copy of the network stream created during the investigation should be stored as part of this digital evidence bag, as outlined in detail by Scanlon et al. \cite{scanlon2014digital}.

%

\section{Proof of Concept}
\label{proofofconcept}

In order to begin proof of concept testing for the investigation methodology, a bespoke BTSync crawling application was first designed and developed. This application was built to emulate regular BTSync client usage, as outlined above, and recorded the necessary results for analysis. 

\subsection{Investigation Overview}
\label{investigationoverview}
To demonstrate the functionality of the application, an investigation was conducted on a known publicly accessible BTSync secret. One of the public BTSync online secret sharing sites was used (\texttt{http://www.bitsynckeys.com/}) to acquire a secret likely to have active peers sharing the corresponding content. The secret selected was advertised with the description ``\texttt{45 GB Movie Collection [Movies] [R]}'' and the read-only secret \texttt{BKV273YUFMWILMESLRDVLI5NHMWO3OCS7} was supplied. It is important to note that there is no certainty that the description accurately advertises the content within the share. There is no method of verifying any of the containing shared content until the syncing process begins and temporary files are created in the shared folder. Even at that point, the user can merely see the filenames of the content once the download/synchronisation process has begun.

\subsection{Results}
\label{results}

As part of the peer identification process a number of active peers were returned to the investigative application. These peers were recorded for later analysis. During the first snapshot taken for this investigation, 21 peers were identified as sharing the specific content and 20 were identified on the second. A snapshot accounts for all of the peers identified sharing the specific content at the same instance in time.

\begin{figure}[t]
\centering
\includegraphics[width=0.5\textwidth]{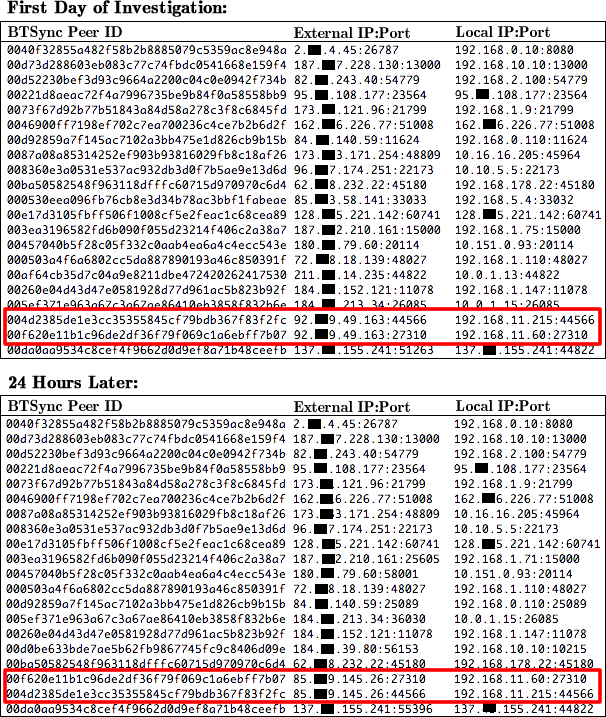}
\caption{Daily Snapshot Comparison for Investigated Secret (Public IP Addresses Partially Redacted)}
\label{comparison}
\end{figure}

Two peers (differentiated by PeerID) of particular interest are listed as the second and third last peers in both tables in Figure \ref{comparison} (highlighted in red). Comparing their peer ID and local IP:Port address pairing, it is clear that these two peers are referring to the same individual node. Between the two snapshots taken of this shared content, their IP address changed from one IP address range to another. However, both of these IP address ranges are associated with the ISP ``Telefonica'' in the same postal zip code in Berlin, Germany (data gathered from Maxmind \cite{maxmind}). This information indicates an ISP level IP address reallocation sometime between the two snapshots as opposed to the use of a VPN or other IP address masking system. The two peers share the same external IP address but have different external ports and local IP:port pairs indicating that the BTSync install on these nodes are accessing the Internet through a router employing Network Address Translation (NAT).

\subsection{Churn Rate}
\label{churn}
While the example investigation outlined as part of this paper focuses on a single secret over a 24 hour window, the low churn rate of just 7\% remains interesting. Most P2P networks experience a high turnover of peers \cite{herrera2007modeling}; following the assumption that most users are active on the network while downloading some content and disconnect upon completion. BTSync is designed to be a tool that functions in a similar manner to cloud file synchronisation services like Dropbox or Google Drive. These tools largely operate on an ``install and forget'' approach whereby synchronisation and updating between the cloud and potentially multiple client machines does not require any direct user input. BTSync uses a similar approach and as a result, low churn rates would be expected.

\subsection{Geolocation}
\label{geolocation}
\begin{figure}
\centering
\includegraphics[width=0.5\textwidth]{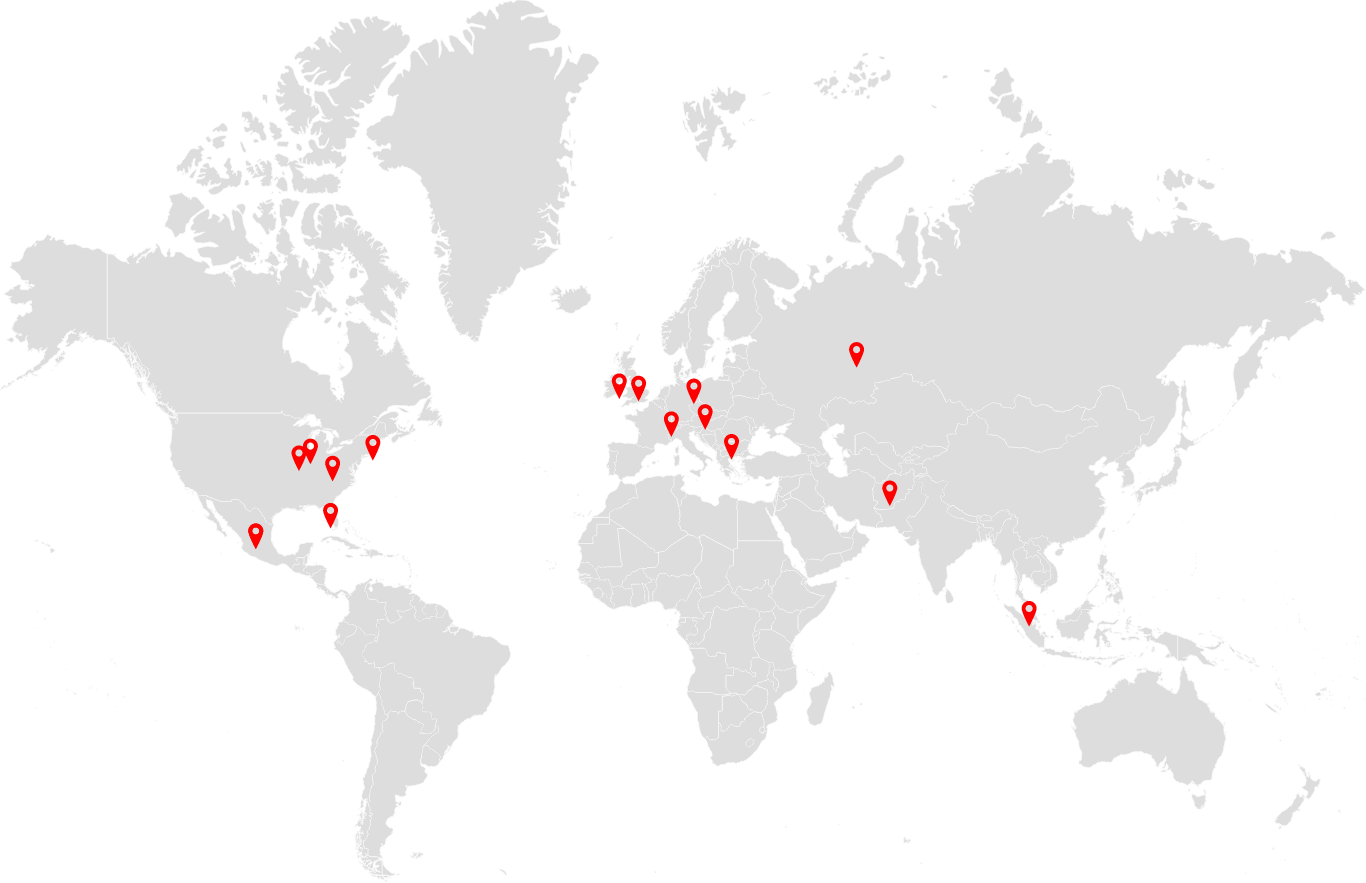}
\caption{Geolocation of Discovered IP Addresses}
\label{fig:geolocation}
\end{figure}

Figure \ref{fig:geolocation} shows the geographic distribution of the peers identified as part of the investigation. While the total number of peers identified with this proof of concept investigation is quite low, the data remains consistent with regular BitTorrent investigative results \cite{scanlon2010week} with North America and Europe being the most popular continents involved.

\section{Conclusion}
\label{conclusion}
This paper documented the protocol used in BitTorrent Sync during the discovery of peers and the synchronisation of data. While BTSync is not necessarily intended to replace BitTorrent as a file dissemination utility, it will likely be used for this purpose. This is already facilitated though websites providing shared secrets, e.g., Reddit \cite{reddit}, etc., as a form of dead-drop. The developers describe the tool as an end-to-end encrypted method of transferring files without the use of a third party staging area, which ensures that the content and personal details remain hidden from unauthorised access. Analysis of the network communication procedure produced unique identifiable information on peers including their unique PeerID, their external and local IP addresses and port numbers. In combination with traditional digital forensic methods, once a secret is identified, it is possible to discover other nodes on the network who are also sharing this data. Deleted data from a local shared folder could be downloaded from the network and recombined for forensic investigation. From an investigative perspective, the decentralised nature of BTSync will always leave an avenue of gathering information and identifying nodes sharing particular content open to the forensic investigator.

\bibliographystyle{IEEEtran}
\bibliography{IEEEabrv,bibfile}

\begin{thebibliography}{10}
\providecommand{\url}[1]{#1}
\csname url@samestyle\endcsname
\providecommand{\newblock}{\relax}
\providecommand{\bibinfo}[2]{#2}
\providecommand{\BIBentrySTDinterwordspacing}{\spaceskip=0pt\relax}
\providecommand{\BIBentryALTinterwordstretchfactor}{4}
\providecommand{\BIBentryALTinterwordspacing}{\spaceskip=\fontdimen2\font plus
\BIBentryALTinterwordstretchfactor\fontdimen3\font minus
  \fontdimen4\font\relax}
\providecommand{\BIBforeignlanguage}[2]{{%
\expandafter\ifx\csname l@#1\endcsname\relax
\typeout{** WARNING: IEEEtran.bst: No hyphenation pattern has been}%
\typeout{** loaded for the language `#1'. Using the pattern for}%
\typeout{** the default language instead.}%
\else
\language=\csname l@#1\endcsname
\fi
#2}}
\providecommand{\BIBdecl}{\relax}
\BIBdecl

\bibitem{bitsync}
\BIBentryALTinterwordspacing
{BitTorrent Inc.} (2013) Bittorrent sync user manual. [Online]. Available:
  \url{http://www.bittorrent.com/help/manual/}
\BIBentrySTDinterwordspacing

\bibitem{bitsyncapi}
\BIBentryALTinterwordspacing
------. (2013) {{BitTorrent Sync Developer API}}. [Online]. Available:
  \url{http://www.bittorrent.com/sync/developers/api}
\BIBentrySTDinterwordspacing

\bibitem{bitsyncstats}
\BIBentryALTinterwordspacing
------. (2013) {{BitTorrent Sync Article}}. [Online]. Available:
  \url{http://blog.bittorrent.com/2013/12/05/
  bittorrent-sync-hits-2-million-user-mark/}
\BIBentrySTDinterwordspacing

\bibitem{cohen2008bittorrent}
\BIBentryALTinterwordspacing
B.~Cohen. (2008) {The BitTorrent Protocol Specification}. [Online]. Available:
  \url{http://bittorrent.org/beps/bep_0003.html/}
\BIBentrySTDinterwordspacing

\bibitem{cohen2003incentives}
------, ``Incentives build robustness in bittorrent,'' in \emph{Proceedings of
  the Workshop on Economics of Peer-to-Peer systems}, vol.~6, 2003, pp. 68--72.

\bibitem{li2005comparing}
J.~Li, J.~Stribling, T.~M. Gil, R.~Morris, and M.~F. Kaashoek, ``{Comparing the
  Performance of Distributed Hash Tables under Churn},'' in \emph{Peer-to-Peer
  Systems III}.\hskip 1em plus 0.5em minus 0.4em\relax Springer, 2005, pp.
  87--99.

\bibitem{sit2002security}
E.~Sit and R.~Morris, ``{Security Considerations for Peer-to-Peer Distributed
  Hash Tables},'' in \emph{Peer-to-Peer Systems}.\hskip 1em plus 0.5em minus
  0.4em\relax Springer, 2002, pp. 261--269.

\bibitem{farina2014}
J.~Farina, M.~Scanlon, and M.-T. Kechadi, ``{BitTorrent Sync: First Impressions
  and Forensic Implications},'' in \emph{{Digital Forensic Research Workshop EU
  (DFRWS EU 2014)}}, May 2014.

\bibitem{layton2010investigation}
R.~Layton and P.~Watters, ``{Investigation into the extent of infringing
  content on BitTorrent networks},'' \emph{Internet Commerce Security
  Laboratory}, 2010.

\bibitem{scanlon2010week}
M.~Scanlon, A.~Hannaway, and M.-T. Kechadi, ``{A Week in the Life of the Most
  Popular BitTorrent Swarms},'' \emph{5th Annual Symposium on Information
  Assurance (ASIA'10)}, 2010.

\bibitem{le2010spying}
S.~Le~Blond, A.~Legout, F.~Lefessant, W.~Dabbous, and M.~A. Kaafar, ``{Spying
  the World from your Laptop: Identifying and Profiling Content Providers and
  Big Downloaders in BitTorrent},'' in \emph{Proceedings of the 3rd USENIX
  conference on Large-scale exploits and emergent threats: botnets, spyware,
  worms, and more}.\hskip 1em plus 0.5em minus 0.4em\relax USENIX Association,
  2010, pp. 4--4.

\bibitem{Chung201281}
H.~Chung, J.~Park, S.~Lee, and C.~Kang, ``{Digital Forensic Investigation of
  Cloud Storage Services},'' \emph{Digital Investigation}, vol.~9, no.~2, pp.
  81 -- 95, 2012.

\bibitem{Martini2013287}
B.~Martini and K.-K.~R. Choo, ``{Cloud storage forensics: ownCloud as a case
  study},'' \emph{Digital Investigation}, vol.~10, no.~4, pp. 287 -- 299, 2013.

\bibitem{reddit}
Reddit. (2014) Btsecrets. http://www.reddit.com/r/btsecrets.

\bibitem{scanlon2014digital}
M.~Scanlon and T.~Kechadi, ``{Digital Evidence Bag Selection for P2P Network
  Investigation},'' in \emph{Future Information Technology}.\hskip 1em plus
  0.5em minus 0.4em\relax Springer, 2014, pp. 307--314.

\bibitem{maxmind}
\BIBentryALTinterwordspacing
M.~Inc. (2014, Jul.) Geolite country database. [Online]. Available:
  \url{http://www.maxmind.com}
\BIBentrySTDinterwordspacing

\bibitem{herrera2007modeling}
O.~Herrera and T.~Znati, ``{Modeling churn in P2P networks},'' in
  \emph{Simulation Symposium, 2007. ANSS'07. 40th Annual}.\hskip 1em plus 0.5em
  minus 0.4em\relax IEEE, 2007, pp. 33--40.

\end{thebibliography}
%



\end{document}